\documentclass[aps,prl,twocolumn,groupedaddress]{revtex4}

\usepackage[utf8]{inputenc}
\usepackage[]{amsmath}
\usepackage[]{graphicx}
\usepackage{amssymb}
\usepackage{float}
\usepackage{bbm}

\newcommand{\de}{\mathrm{d}}

\newcommand{\ddt}{\frac{\mathrm{d}}{\mathrm{d}t}}
\newcommand{\De}{\mathrm{D}}
\newcommand{\obs}{\mathrm{obs}}

\renewcommand{\mathbb}[1]{\mathbbm{#1}} 

\begin{document}


\title{A Unified Approach to Integration and Optimization of Parametric Ordinary Differential Equations}


\author{Daniel Kaschek}
\affiliation{Institute of Physics, Freiburg University, Freiburg, Germany}
\author{Jens Timmer}
\affiliation{Institute of Physics, Freiburg University, Freiburg, Germany\\
	BIOSS Centre for Biological Signalling Studies, Freiburg University, Germany\\
	Freiburg Center for Systems Biology (ZBSA), Freiburg University, Freiburg, Germany\\
	Freiburg Institute for Advanced Studies (FRIAS), Freiburg University, Freiburg, Germany
}

\date{\today}


\begin{abstract}
Parameter estimation in ordinary differential equations, although applied and refined in various fields of the quantitative sciences, is still confronted with a variety of difficulties. One major challenge is finding the global optimum of a log-likelihood function that has several local optima, e.g.~in oscillatory systems. In this publication, we introduce a formulation based on continuation of the log-likelihood function that allows to restate the parameter estimation problem as a boundary value problem. By construction, the ordinary differential equations are solved and the parameters are estimated both in one step. The formulation as a boundary value problem enables an optimal transfer of information given by the measurement time courses to the solution of the estimation problem, thus favoring convergence to the global optimum. This is demonstrated explicitly for the fully as well as the partially observed Lotka-Volterra system.

\end{abstract}

\maketitle

Ordinary differential equation (ODE) models play a key role for understanding and predicting the behavior of dynamic systems originating from various disciplines like physics, chemistry or the life sciences. In many cases, these dynamic models depend on parameters that are not known beforehand but need to be determined from measurement data by means of statistical methods. Inference of parameters of dynamic systems from measurement data is commonly realized by optimization of the likelihood function. Optimization is a broad field and many different algorithms have come up over the last decades \cite{parlitz_estimating_1996, sohl-dickstein_new_2011, amritkar_estimating_2009, sitz_estimation_2002, horbelt_identifying_2001}, each of them with problem specific advantages and disadvantages. One characteristic distinction between optimizers is whether they include stochasticity or not. Stochastic optimizers, e.g.~evolutionary algorithms \cite{goswami_learning_2007}, particle swarms \cite{mendes_fully_2004, peng_parameter_2010} or simulated annealing \cite{xiang_efficiency_2000} are especially valuable for discontinuous likelihood functions where gradient information is not available or not defined. On the other hand, many deterministic optimizers employ information about the differentiable structure of the likelihood, i.e.~gradient and Hessian information. For differentiable likelihood functions this has the advantage that convergence is achieved much faster. However, this approach has to struggle with other difficulties. If the likelihood function has local optima, the outcome of the optimization procedure depends on the starting point. Once the optimizer is approaching a local optimum, the algorithm will not leave this optimum disregarding the existence of better optima.\\
The problem of local optima has been addressed by several approaches. It has been shown that a combination of deterministic and stochastic optimization can help escaping local optima and finding the global optimum \cite{villaverde_cooperative_2012}. Other approaches modify the dynamic system by homotopy transformations \cite{vyasarayani_single-shooting_2012} introducing a factor~$\lambda$ that allows for a continuous transition between the modified, convex problem and the original problem. Hence another approach is the multiple-shooting method \cite{bock_numerical_2007}. Most optimizers follow a single-shooting approach, i.e.~model trajectories are computed based on given initial values and the outcome is compared to the data. In contrast, the multiple-shooting approach introduces a grid of time-points and initial condition parameters. The optimizer is initialized with discontinuous trajectories and constraints are defined guaranteeing that all trajectories become continuous in the course of optimization.\\
In our work, we present a reformulation of the optimization problem as a boundary value problem (BVP). The motivation for this approach is two-fold. The first argument follows from the history of gradient-based single-shooting optimization for parameter estimation in ordinary differential equations. The performance and accuracy of this method has been enormously increased by solving the ODE together with its' derivatives with respect to the parameters, i.e.~the sensitivity equations, in one integration run. This augmentation step allows a fast and accurate computation of the gradient but still evaluation and optimization of the objective function are separate steps. Our aim is to take the next logical step and incorporate even optimization into the ODE integration. The second argument takes up the multiple-shooting idea: the possibility to initialize BVP solvers with prior knowledge like approximate trajectories from measurement data. If the optimization problem is equivalently expressed as a boundary value problem then a good initialization should increase the solver's ability to find the correct solution.\\
In the following, we show how both objectives can be matched. Our augmentation of the ODE is based on continuation of the log-likelihood function to a differentiable function of time. The resulting system constitutes a BVP. By construction, the solution of this BVP is optimal with respect to the log-likelihood function and it can be obtained by standard numerical BVP solvers. The initialization of the BVP solver allows for an efficient transfer of information provided by the observation data.\\

We consider a dynamic system defined by ordinary differential equations (ODE),
\begin{align}
	\ddt x = f(x, p),\quad x(0)=x_0,
	\label{eq:dynSystem}
\end{align}
with time $t$, states $x\in\mathbb R^n$ and parameters $p\in\mathbb R^r$. The extension of the system by $\ddt p = 0$ transforms the parameters into usual state variables. For the augmented states $\xi = (x, p)$, parameter estimation becomes an estimation of initial conditions. Furthermore, let $x_{\obs} = (x_1, \dots, x_m)$, with $m\leq n$, be the observed states and let $\{x_1^D(t_j), \dots, x_m^D(t_j)\}_j$ denote the time-discrete observation data. The observation data can be approximated by a continuous data function $x_{\obs}^D(t) = (x_1^D(t), \dots, x_m^{D}(t))$, e.g.~by linear interpolation or spline interpolation. On the other hand, we assume that measurement events for different time points are statistically independent, consequently, the likelihood function
\begin{align}
	L(\xi_0 | \{x_{\obs}^D(t_j)\}_j) = \prod_j L_j(\xi_0 | x_{\obs}^D(t_j))
	\label{}
\end{align}
factorizes and the negative log-likelihood
\begin{align}
	\ell(\xi_0 | \{x_{\obs}^D(t_j)\}_j) &= \sum_j -\log L_j(\xi_0, x_{\obs}^D(t_j))\\
		&\approx \frac{1}{T}\int_0^T r(\xi_0, t)\de t
	\label{}
\end{align}
can be approximated by the integral. Here, $r(\xi_0,t)$ denotes the continuous approximation of $-\log L_j(\xi_0, x_{\obs}^D(t_j))$. For standard normally distributed noise, $r(\xi_0,t)$ becomes $\big(x_{\obs}(t)-x_{\obs}^D(t)\big)^2$ which will be used in the following. The argumentation also holds for other noise distributions.\\
An initial condition vector $\hat{\xi}_0=(\hat{x}_0, \hat{p}_0)\in\mathbb R^{n+r}$ is a local optimum if $\nabla \ell(\hat{\xi}_0, t=T) = 0$ vanishes at the latest observed time point $T$. Since $\ell(\xi_0, t=0) = 0$ for all values of $\xi_0$ at initial time, the gradient $\nabla \ell(\hat{\xi}_0, t=0) = 0$ vanishes, too. This observation constitutes the boundary condition that needs to be matched for a local optimum, i.e.
\begin{align}
	\nabla \ell(\hat{\xi}_0, 0) = \nabla \ell(\hat{\xi}_0, T) = 0.
	\label{eq:bc}
\end{align}
Each line of eq.~\eqref{eq:bc} has the potential to determine one parameter value. In order to include this condition into the dynamic system \eqref{eq:dynSystem}, $\nabla \ell$ is derived with respect to time. At this point it is crucial having approximated the negative log-likelihood by an integral expression:
\begin{align}
	\ddt \nabla \ell&(\xi_0 | x_{\obs}^D(t)) =  \frac{1}{T}\ddt \int_0^t \nabla r(\xi_0, \tau)\de\tau\\
	&=\frac{1}{T} \nabla r(\xi_0, t)\\
	&= \frac{2}{T}\big(x_{\obs}(t) - x_{\obs}^D(t)\big)^{*}\De_{\xi_0} x_{\obs}(t).
	\label{}
\end{align}
Here, `` $^*$ '' indicates the transpose and $\De_{\xi_0} x_{\obs}(t)$ denotes the Jacobian of $x_{\obs}(t)$ with respect to the initial conditions $\xi_0$, also known as the sensitivities of the solution trajectory $x_{\obs}(t)$. The sensitivities are determined by an ODE, too, hence the complete systems reads
\begin{align}
	\ddt \xi &= f(\xi) \label{eq:ode}\\
	\ddt \De_{\xi_0}\xi &= \De_{\xi}f\, \De_{\xi_0}\xi \label{eq:sens}\\
	\ddt \nabla \ell &= \frac{2}{T}\big(x_{\obs} - x_{\obs}^D\big)^{*}\,\De_{\xi_0} x_{\obs}. \label{eq:gradient}
\end{align}
The sensitivity equations \eqref{eq:sens} have fixed initial conditions, $\textrm{diag}(\mathbb 1_{n+r})$, with the identity matrix $\mathbb 1_{n+r}\in\mathbb R^{(n+r)\times (n+r)}$. The gradient equations \eqref{eq:gradient} have both zero initial \textit{and} final condition, see eq.~\eqref{eq:bc}, a boundary constraint that fully determines the initial values of the augmented states in eq.~\eqref{eq:ode}. On the other hand, these are the parameters and initial conditions we seek to estimate. Hence, the desired values $\hat{\xi}_0$ optimizing the negative log-likelihood are part of the solution of the two-point boundary value problem, eqs.~(\ref{eq:ode}-\ref{eq:gradient}). This is the principle behind our optimization approach.\\ 
The solution of this two-point boundary value problem can be numerically obtained by different solver implementations, see \cite{simos_efficient_2011} for a survey. Compared to gradient based single-shooting methods, eq.~\eqref{eq:gradient} represents the pivotal difference. It translates optimization into the ambit of integration.\\

\begin{figure}[H]
	\begin{center}
	\includegraphics[width=0.39\textwidth]{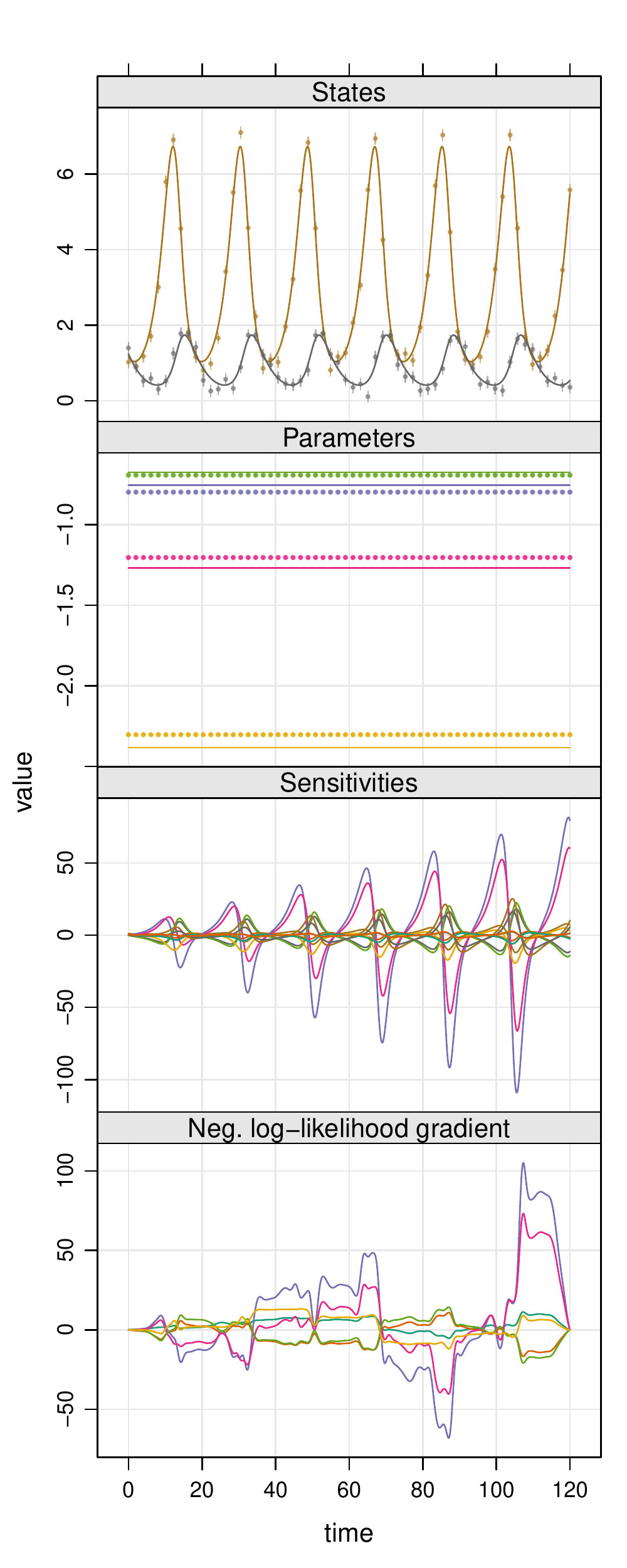}
	\end{center}
	\caption{Solution generated by the BVP solver. The 4 panels show state solutions with simulated data points, parameter solutions on a logarithmic scale with true values as dots, sensitivity solutions and the gradient of the negative log-likelihood.}
	\label{fig:bvpsolution}
\end{figure}

In the following, we examine the Lotka-Volterra equations \cite{lotka_1909, wang_population_2012, parker_extinction_2009}. They give a basic description of the predator and prey population dynamics. The system is defined by two differential equations
\begin{align}
	\ddt A &= A(\alpha - \beta B)\label{eq:lv1},\\
	\ddt B &= -B(\gamma -\delta A)\label{eq:lv2},
\end{align}
where $A$ and $B$ correspond to prey and predator, respectively. The parameters $\alpha$ and $\beta$ describe prey reproduction and reduction, $\gamma$ and $\delta$ describe predator extinction and reproduction. For non-zero initial condition, the solution of eqs.~(\ref{eq:lv1}-\ref{eq:lv2}) is a sustained oscillation.\\
In the first part of the study, it is assumed that both populations are observed. Observation data is simulated by numerically integrating the ODE system with $A_0 = 1$, $B_0 = 1.2$, $\alpha = 0.45$, $\beta = 0.5$, $\gamma = 0.3$, $\delta = 0.1$ and adding Gaussian noise with $\sigma=0.15$ to the solution. Subsequently, the parameter values are sought to be recovered from randomly chosen initial parameter guesses. The initialization of the BVP solver, i.e.~the grid of initially assumed values for each of the variables in eqs.~(\ref{eq:ode}-\ref{eq:gradient}), is obtained by integrating the sensitivity equations \eqref{eq:sens} with the initial parameter guess and the data interpolations as input trajectories. The gradient is assumed to vanish over the entire range.\\
Independently of the initial parameter values, the BVP solver converges to the same solution. The result for one representative data set is shown in Figure~\ref{fig:bvpsolution}. The States panel shows the solutions of the state variables $A$ and $B$ together with the data points and error bars. As expected, the predator and prey trajectories hit about 67\% of the error bars. In the Parameters panel, the solutions of the state variables $\alpha$, $\beta$, $\gamma$ and $\delta$ are shown on a logarithmic scale, i.e. the dynamic parameters. The dots indicate the values that have been used for simulation. The Sensitivities panel shows the sensitivity trajectories which are typical for oscillating systems, i.e.~oscillations with increasing amplitude. Finally, in the Negative log-likelihood gradient panel, the gradient solution is plotted. The time scale of gradient changes is determined by the sampling density of the simulated time course. It hits the ground line in the end point as desired, guaranteeing an optimum.\\

\begin{figure}[H]
	\includegraphics[width=0.47\textwidth, page=1]{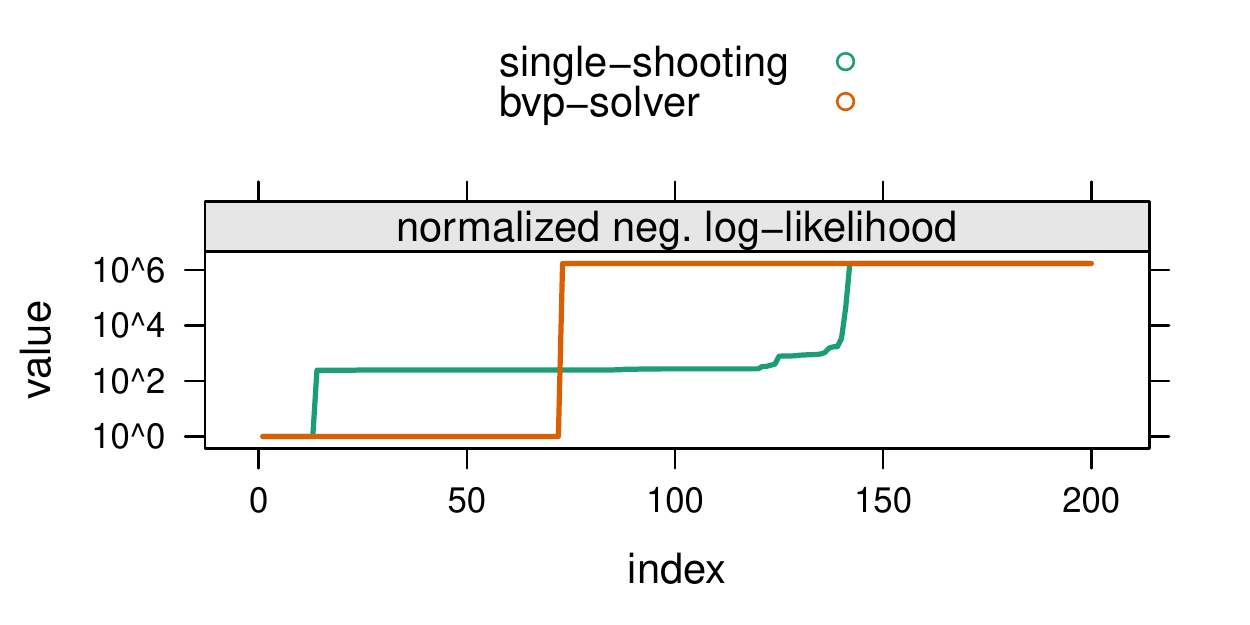}\\
	\includegraphics[width=0.49\textwidth, page=1]{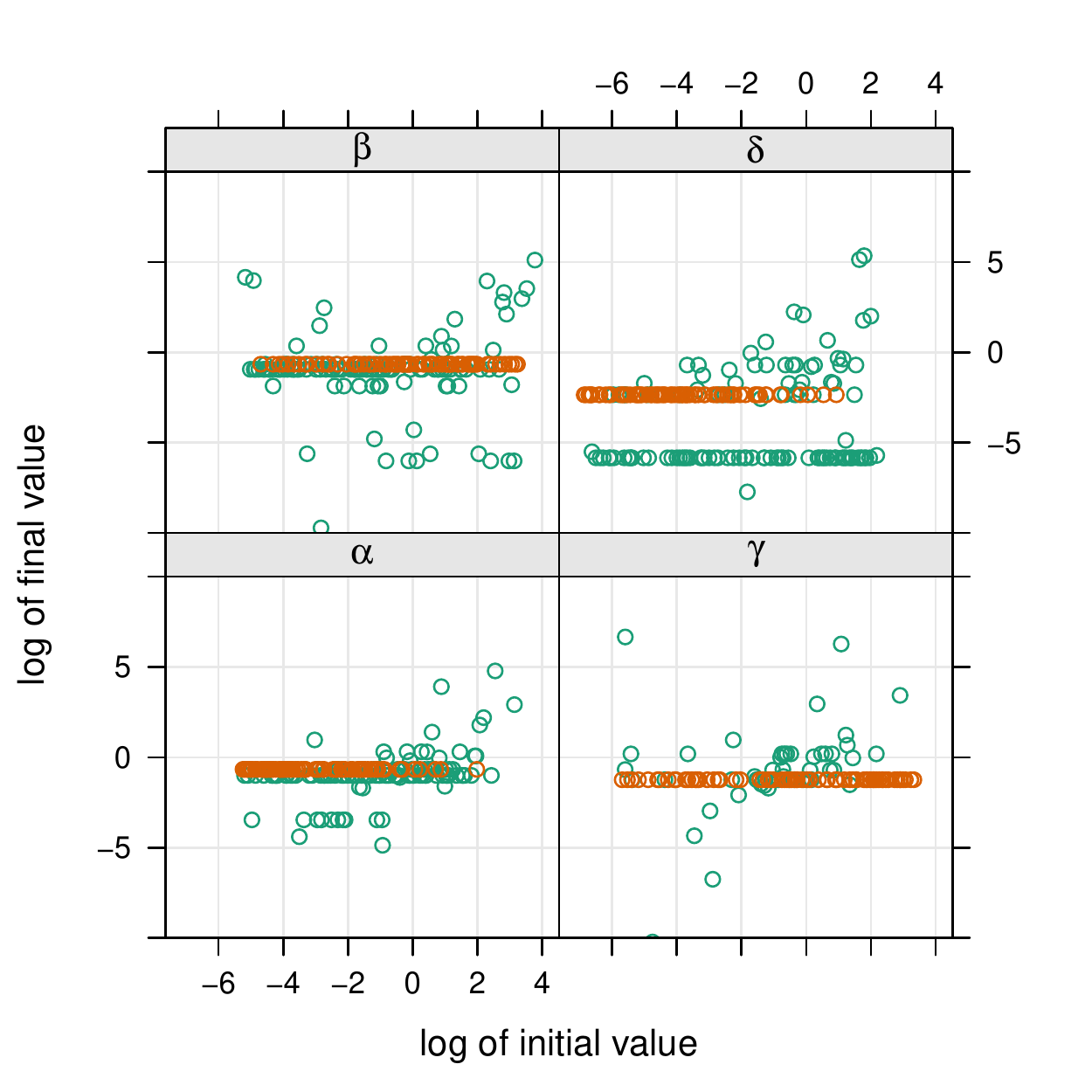}
	\caption{Comparison of single-shooting and BVP method tested on the fully observed Lotka-Volterra system. Each method has been applied to simulated data sets and for each data set, initial parameter vectors have been generated by Latin hypercube sampling covering a range of 4 orders of magnitude around the true parameter values. The resulting negative log-likelihood values were sorted, normalized by the smallest value and plotted on a logarithmic scale. In the scatter plots, initial parameter vectors are plotted against final parameter values for each optimization resulting in a negative log-likelihood value lower than $10^3$.}
	\label{fig:stairway}
\end{figure}

The BVP method has been tested systematically against a single-shooting approach based on the Levenberg-Marquardt algorithm. For different simulated data sets, both, BVP method and single-shooting method have been applied to the same random set of initial parameter guesses. In order to avoid that, by chance, parameter vectors are too similar, we employed Latin hypercube sampling with a hypercube covering 4 orders of magnitude around the true parameter values. Both optimization approaches failed convergence a number of times in which case $10^6$ was assigned as value to the negative log-likelihood. Figure~\ref{fig:stairway} shows the first 200 sorted negative log-likelihood values of the total 300 initial guesses for both approaches.

The single-shooting approach gets stuck in different local optima and finds the global optimum only in 4\% of the cases. In contrast, the BVP method proves to be robust against different initial guesses for the parameter values. The solution converges either to the global minimum or it fails convergence. It is 6 times more efficient in finding the best optimum. Figure~\ref{fig:stairway} also gives some indication about parameter convergence regions. For the BVP method, the set of initial parameters that finally converged to the best parameter value covers almost the total range. However, fewer initial guesses with $\alpha$ and $\delta$ larger than 1 lead to a successful reconstruction of the BVP solution. The broad plateau of local optima for the single-shooting method is reflected in a clear shift of final parameter values and a certain number of randomly distributed final parameters.

\begin{figure}[H]
	\includegraphics[width=0.46\textwidth, page=2]{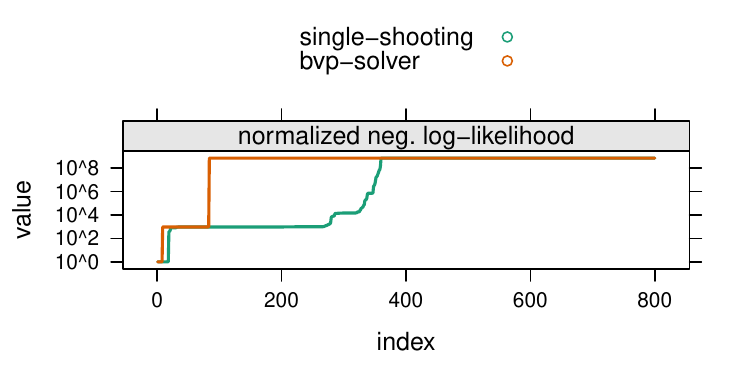}\\
	\includegraphics[width=0.48\textwidth, page=2]{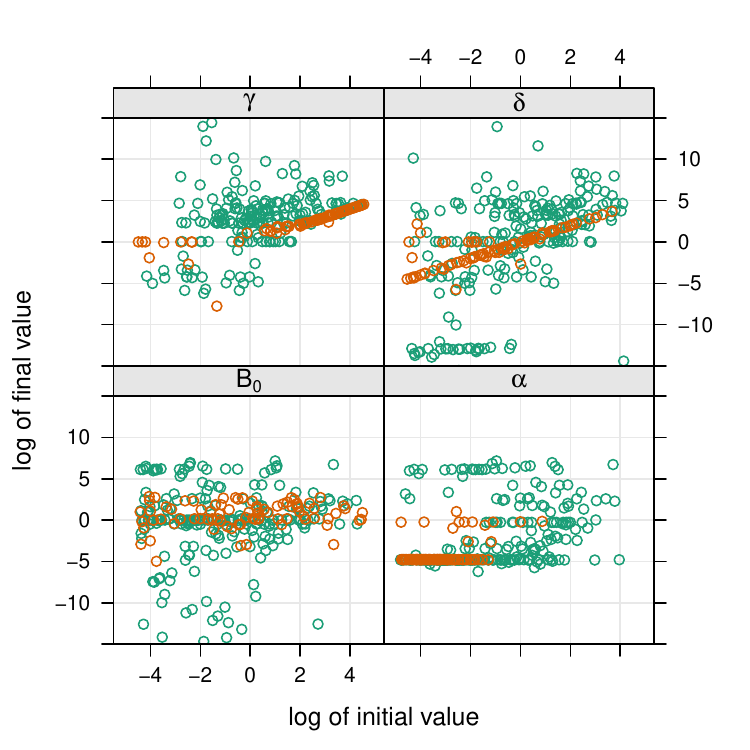}
	\caption{Comparison of single-shooting and BVP method tested on the partially observed Lotka-Volterra system. Each method has been applied to simulated data sets and for each data set, initial parameter vectors have been generated by Latin hypercube sampling covering a range of 4 orders of magnitude around the true parameter values. The resulting negative log-likelihood values were sorted, normalized by the smallest value and plotted on a logarithmic scale. In the scatter plots, initial parameter vectors are plotted against final parameter values for each optimization resulting in a negative log-likelihood value lower than $10^3$.}
	\label{fig:stairway2}
\end{figure}

In a second step, the observation of $B$ is omitted and~$\beta$ is fixed to 1 in order to keep the system identifiable.
Analogously to the fully observed system, data sets have been simulated and Gaussian noise has been added. The comparison between the single-shooting method and the BVP method is shown in Figure \ref{fig:stairway2}. The plots indicate that the situation becomes more intricate if only one state is observed. The convergence rate drops below 2\% for both approaches and the BVP method reconstructs a variety of local optima, each of them with an almost identical negative log-likelihood value. From the scatter plots in Figure \ref{fig:stairway2}, two conclusions can be drawn for the BVP method: First, each of the globally convergent solutions started from the negative orthant and second, there is a submanifold with boundary in the parameter space corresponding to the same locally optimal negative log-likelihood value.

\begin{figure}[H]
	\includegraphics[width=0.46\textwidth, page=3]{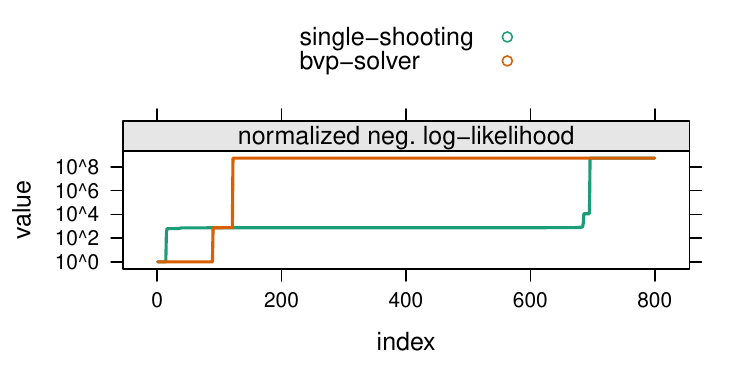}\\
	\includegraphics[width=0.48\textwidth, page=3]{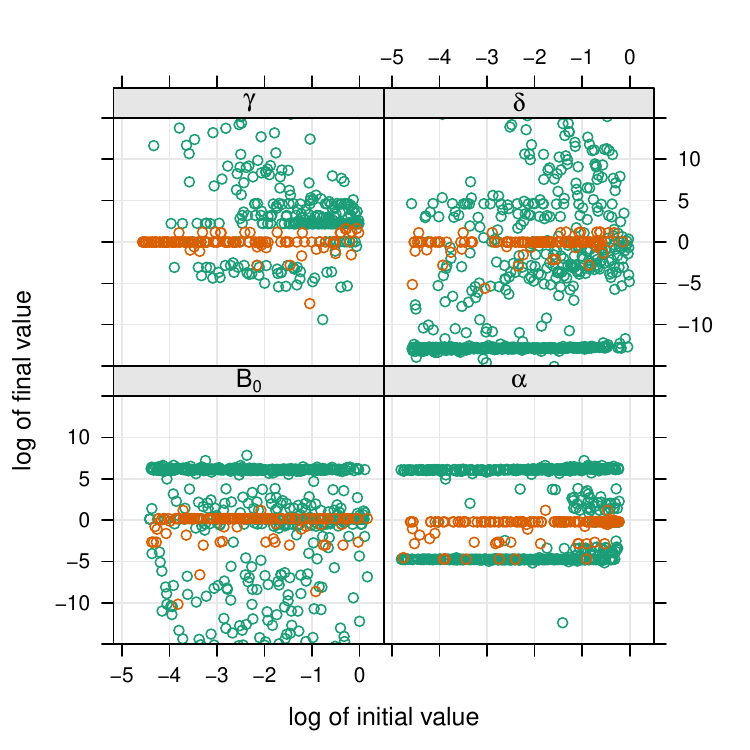}
	\caption{Comparison of single-shooting and BVP method tested on the partially observed Lotka-Volterra system. Each method has been applied to simulated data sets and for each data set, initial parameter vectors have been generated by Latin hypercube sampling covering the negative orthant of parameter space. The resulting negative log-likelihood values were sorted, normalized by the smallest value and plotted on a logarithmic scale. In the scatter plots, initial parameter vectors are plotted against final parameter values for each optimization resulting in a negative log-likelihood value lower than $10^3$.}
	\label{fig:stairway3}
\end{figure}

Figure \ref{fig:stairway3} shows the same picture for initial guesses starting from the negative orthant only. In agreement with the expectation, the number of BVP solutions corresponding to the global optimum increases considerably and exceeds the success rate of the single-shooting method by a factor of 5.
Finally, we come to the following conclusions: In case of a partially observed system, the success rate of the BVP method depends on favorable initial conditions. Whereas the single-shooting algorithm performed equally badly over the entire parameter space, for the boundary value approach, it was possible to identify an attractive basin resulting in a considerably increased convergence rate.\\
From the fully observed Lotka-Volterra system, we conclude that the boundary value approach is excellently suited as optimization approach if numerous observables are available. In this case, it clearly outperforms the single-shooting Levenberg-Marquardt algorithm in terms of convergence to the global optimum. The strength of the presented optimization approach is its ability to exploit the measured time courses in a natural way. This favors convergence to the global optimum. Unlike single-shooting approaches, convergence to local optima or false convergence claims are efficiently reduced.\\
The key to this advantage is the reformulation of the estimation problem as a boundary value problem which, in turn, is enabled by continuation of the negative log-likelihood function to a time-differentiable function. This restatement elegantly incorporates optimization and ODE solution in one task. By nature of the boundary value problem, an initial guess for all state variables needs to be presented to the numerical solver. The initialization by measured time-courses carries exactly the information that is necessary to make the algorithm converge to the best optimum.\\
This work was supported by the MIP-DILI project, Innovative Medicines Initiative Joint Undertaking under grant agreement No.~115336. We thank our colleague Marcus Rosenblatt for supporting the computational implementation.

\bibliography{bvpestimation2}

\begin{thebibliography}{16}
\expandafter\ifx\csname natexlab\endcsname\relax\def\natexlab#1{#1}\fi
\expandafter\ifx\csname bibnamefont\endcsname\relax
  \def\bibnamefont#1{#1}\fi
\expandafter\ifx\csname bibfnamefont\endcsname\relax
  \def\bibfnamefont#1{#1}\fi
\expandafter\ifx\csname citenamefont\endcsname\relax
  \def\citenamefont#1{#1}\fi
\expandafter\ifx\csname url\endcsname\relax
  \def\url#1{\texttt{#1}}\fi
\expandafter\ifx\csname urlprefix\endcsname\relax\def\urlprefix{URL }\fi
\providecommand{\bibinfo}[2]{#2}
\providecommand{\eprint}[2][]{\url{#2}}

\bibitem[{\citenamefont{Parlitz}(1996)}]{parlitz_estimating_1996}
\bibinfo{author}{\bibfnamefont{U.}~\bibnamefont{Parlitz}},
  \bibinfo{journal}{Physical Review Letters} \textbf{\bibinfo{volume}{76}},
  \bibinfo{pages}{1232} (\bibinfo{year}{1996}).

\bibitem[{\citenamefont{Sohl-Dickstein
  et~al.}(2011)\citenamefont{Sohl-Dickstein, Battaglino, and
  {DeWeese}}}]{sohl-dickstein_new_2011}
\bibinfo{author}{\bibfnamefont{J.}~\bibnamefont{Sohl-Dickstein}},
  \bibinfo{author}{\bibfnamefont{P.~B.} \bibnamefont{Battaglino}},
  \bibnamefont{and} \bibinfo{author}{\bibfnamefont{M.~R.}
  \bibnamefont{{DeWeese}}}, \bibinfo{journal}{Physical Review Letters}
  \textbf{\bibinfo{volume}{107}}, \bibinfo{pages}{220601}
  (\bibinfo{year}{2011}).

\bibitem[{\citenamefont{Amritkar}(2009)}]{amritkar_estimating_2009}
\bibinfo{author}{\bibfnamefont{R.~E.} \bibnamefont{Amritkar}},
  \bibinfo{journal}{Physical Review E} \textbf{\bibinfo{volume}{80}},
  \bibinfo{pages}{047202} (\bibinfo{year}{2009}).

\bibitem[{\citenamefont{Sitz et~al.}(2002)\citenamefont{Sitz, Schwarz, Kurths,
  and Voss}}]{sitz_estimation_2002}
\bibinfo{author}{\bibfnamefont{A.}~\bibnamefont{Sitz}},
  \bibinfo{author}{\bibfnamefont{U.}~\bibnamefont{Schwarz}},
  \bibinfo{author}{\bibfnamefont{J.}~\bibnamefont{Kurths}}, \bibnamefont{and}
  \bibinfo{author}{\bibfnamefont{H.~U.} \bibnamefont{Voss}},
  \bibinfo{journal}{Physical Review E} \textbf{\bibinfo{volume}{66}},
  \bibinfo{pages}{016210} (\bibinfo{year}{2002}).

\bibitem[{\citenamefont{Horbelt et~al.}(2001)\citenamefont{Horbelt, Timmer,
  J.~Bunner, Meucci, and Ciofini}}]{horbelt_identifying_2001}
\bibinfo{author}{\bibfnamefont{W.}~\bibnamefont{Horbelt}},
  \bibinfo{author}{\bibfnamefont{J.}~\bibnamefont{Timmer}},
  \bibinfo{author}{\bibfnamefont{M.}~\bibnamefont{J.~Bunner}},
  \bibinfo{author}{\bibfnamefont{R.}~\bibnamefont{Meucci}}, \bibnamefont{and}
  \bibinfo{author}{\bibfnamefont{M.}~\bibnamefont{Ciofini}},
  \bibinfo{journal}{Physical Review E} \textbf{\bibinfo{volume}{64}},
  \bibinfo{pages}{016222} (\bibinfo{year}{2001}).

\bibitem[{\citenamefont{Goswami and Liu}(2007)}]{goswami_learning_2007}
\bibinfo{author}{\bibfnamefont{G.}~\bibnamefont{Goswami}} \bibnamefont{and}
  \bibinfo{author}{\bibfnamefont{J.~S.} \bibnamefont{Liu}},
  \bibinfo{journal}{Statistics and Computing} \textbf{\bibinfo{volume}{17}},
  \bibinfo{pages}{23} (\bibinfo{year}{2007}), ISSN \bibinfo{issn}{0960-3174,
  1573-1375}.

\bibitem[{\citenamefont{Mendes et~al.}(2004)\citenamefont{Mendes, Kennedy, and
  Neves}}]{mendes_fully_2004}
\bibinfo{author}{\bibfnamefont{R.}~\bibnamefont{Mendes}},
  \bibinfo{author}{\bibfnamefont{J.}~\bibnamefont{Kennedy}}, \bibnamefont{and}
  \bibinfo{author}{\bibfnamefont{J.}~\bibnamefont{Neves}},
  \bibinfo{journal}{{IEEE} Transactions on Evolutionary Computation}
  \textbf{\bibinfo{volume}{8}}, \bibinfo{pages}{204 } (\bibinfo{year}{2004}),
  ISSN \bibinfo{issn}{1089-{778X}}.

\bibitem[{\citenamefont{Peng et~al.}(2010)\citenamefont{Peng, Li, Yang, and
  Liu}}]{peng_parameter_2010}
\bibinfo{author}{\bibfnamefont{H.}~\bibnamefont{Peng}},
  \bibinfo{author}{\bibfnamefont{L.}~\bibnamefont{Li}},
  \bibinfo{author}{\bibfnamefont{Y.}~\bibnamefont{Yang}}, \bibnamefont{and}
  \bibinfo{author}{\bibfnamefont{F.}~\bibnamefont{Liu}},
  \bibinfo{journal}{Physical Review E} \textbf{\bibinfo{volume}{81}},
  \bibinfo{pages}{016207} (\bibinfo{year}{2010}).

\bibitem[{\citenamefont{Xiang and Gong}(2000)}]{xiang_efficiency_2000}
\bibinfo{author}{\bibfnamefont{Y.}~\bibnamefont{Xiang}} \bibnamefont{and}
  \bibinfo{author}{\bibfnamefont{X.~G.} \bibnamefont{Gong}},
  \bibinfo{journal}{Physical Review E} \textbf{\bibinfo{volume}{62}},
  \bibinfo{pages}{4473} (\bibinfo{year}{2000}).

\bibitem[{\citenamefont{Villaverde et~al.}(2012)\citenamefont{Villaverde, Egea,
  and Banga}}]{villaverde_cooperative_2012}
\bibinfo{author}{\bibfnamefont{A.~F.} \bibnamefont{Villaverde}},
  \bibinfo{author}{\bibfnamefont{J.~A.} \bibnamefont{Egea}}, \bibnamefont{and}
  \bibinfo{author}{\bibfnamefont{J.~R.} \bibnamefont{Banga}},
  \bibinfo{journal}{{BMC} Systems Biology} \textbf{\bibinfo{volume}{6}},
  \bibinfo{pages}{75} (\bibinfo{year}{2012}), ISSN \bibinfo{issn}{1752-0509}.

\bibitem[{\citenamefont{Vyasarayani et~al.}(2012)\citenamefont{Vyasarayani,
  Uchida, and {McPhee}}}]{vyasarayani_single-shooting_2012}
\bibinfo{author}{\bibfnamefont{C.~P.} \bibnamefont{Vyasarayani}},
  \bibinfo{author}{\bibfnamefont{T.}~\bibnamefont{Uchida}}, \bibnamefont{and}
  \bibinfo{author}{\bibfnamefont{J.}~\bibnamefont{{McPhee}}},
  \bibinfo{journal}{Physical Review E} \textbf{\bibinfo{volume}{85}}
  (\bibinfo{year}{2012}), ISSN \bibinfo{issn}{1539-3755, 1550-2376}.

\bibitem[{\citenamefont{Bock et~al.}(2007)\citenamefont{Bock, Kostina, and
  Schlöder}}]{bock_numerical_2007}
\bibinfo{author}{\bibfnamefont{H.~G.} \bibnamefont{Bock}},
  \bibinfo{author}{\bibfnamefont{E.}~\bibnamefont{Kostina}}, \bibnamefont{and}
  \bibinfo{author}{\bibfnamefont{J.~P.} \bibnamefont{Schlöder}},
  \bibinfo{journal}{{GAMM-Mitteilungen}} \textbf{\bibinfo{volume}{30}},
  \bibinfo{pages}{376–408} (\bibinfo{year}{2007}).

\bibitem[{\citenamefont{Cash and Mazzia}(2011)}]{simos_efficient_2011}
\bibinfo{author}{\bibfnamefont{J.~R.} \bibnamefont{Cash}} \bibnamefont{and}
  \bibinfo{author}{\bibfnamefont{F.}~\bibnamefont{Mazzia}}, in
  \emph{\bibinfo{booktitle}{Recent Advances in Computational and Applied
  Mathematics}}, edited by \bibinfo{editor}{\bibfnamefont{T.~E.}
  \bibnamefont{Simos}} (\bibinfo{publisher}{Springer Netherlands},
  \bibinfo{address}{Dordrecht}, \bibinfo{year}{2011}), pp.
  \bibinfo{pages}{23--39}, ISBN \bibinfo{isbn}{978-90-481-9980-8,
  978-90-481-9981-5}.

\bibitem[{\citenamefont{Lotka}(1909)}]{lotka_1909}
\bibinfo{author}{\bibfnamefont{A.~J.} \bibnamefont{Lotka}},
  \bibinfo{journal}{The Journal of Physical Chemistry}
  \textbf{\bibinfo{volume}{14}}, \bibinfo{pages}{271} (\bibinfo{year}{1909}).

\bibitem[{\citenamefont{Wang and Lai}(2012)}]{wang_population_2012}
\bibinfo{author}{\bibfnamefont{M.-X.} \bibnamefont{Wang}} \bibnamefont{and}
  \bibinfo{author}{\bibfnamefont{P.-Y.} \bibnamefont{Lai}},
  \bibinfo{journal}{Physical Review E} \textbf{\bibinfo{volume}{86}},
  \bibinfo{pages}{051908} (\bibinfo{year}{2012}).

\bibitem[{\citenamefont{Parker and Kamenev}(2009)}]{parker_extinction_2009}
\bibinfo{author}{\bibfnamefont{M.}~\bibnamefont{Parker}} \bibnamefont{and}
  \bibinfo{author}{\bibfnamefont{A.}~\bibnamefont{Kamenev}},
  \bibinfo{journal}{Physical Review E} \textbf{\bibinfo{volume}{80}},
  \bibinfo{pages}{021129} (\bibinfo{year}{2009}).

\end{thebibliography}
\vspace{2cm}

\end{document}